# A pragmatic look at education and training of software test engineers: Further cooperation of academia and industry is needed

Vahid Garousi
Queen's University Belfast, UK
Testinium A.Ş., Türkiye
ProSys MMC, Azerbaijan
v.garousi@qub.ac.uk

Alper Buğra Keleş
Testinium A.Ş., Türkiye
alper.keles@testinium.com

*Abstract*--Alongside software testing education in universities, a great extent of effort and resources are spent on software-testing training activities in industry. For example, there are several international certification schemes in testing, such as those provided by the International Software Testing Qualifications Board (ISTQB), which have been issued to more than 914K testers so far. To train the highly qualified test engineers of tomorrow, it is important for both university educators and trainers in industry to be aware of the status of software testing education in academia versus its training in industry, to analyze the relationships of these two approaches, and to assess ways on how to improve the education / training landscape. For that purpose, this paper provides a pragmatic overview of the issue, presents several recommendations, and hopes to trigger further discussions in the community, between industry and academia, on how to further improve the status-quo, and to find further best practices for more effective education and training of software testers. The paper is based on combined ~40 years of the two authors' technical experience in test engineering, and their ~30 years of experience in providing testing education and training in more than six countries.

*Keywords*--*Software testing; software-testing education; training; education research; experience-based education research*

## 1 INTRODUCTION

With the increasing complexity and scale of software systems, there is an ever-increasing demand for sophisticated and cost-effective software testing and quality assurance practices. To meet such a demand, there is an ever-increasing need for highly-skilled software testing work-force (test engineers) in the industry.

To train highly-skilled software test engineers, many university Software Engineering (SE) or Computer Science (CS) degree programs offer software-testing education, either as separate software testing courses, or by blending testing concepts into various programming and/or software-engineering courses [1].

Alongside software testing education in most universities, a great amount of efforts and resources are spent on software-testing training activities in industry, e.g., there are several certification schemes in testing, such as those provided by the International Software Testing Qualifications Board (ISTQB) (istqb.org), a non-for-profit organization that has national branches in more than 120 countries worldwide. According to the ISTQB website, "*As of June 2023, ISTQB® has administered over 1.2 million exams and issued more than 914k certifications in over 130 countries*". Furthermore, a Google search for "software testing training" returns about 663 million search results, in the grey literature [2], as of this writing (Jan. 2024); thus highlighting the substantial world-wide interest by learners to learn various concepts and skills of software testing.

It is important for both university educators and trainers/consultants in industry to analyze the status of software testing education in academia versus its training in industry, and to know the relationships of these two approaches to train software-testing professionals of the future. The goal of this paper is to provide a pragmatic review of the issue and to trigger further discussions in the community, and between industry and academia, on how to further improve the status quo and also to find further best practices with cooperation of industry and academia in education and training of software testers.

The remainder of this paper is organized as follows. Section 2 reviews the related work. In Section 3, we provide a visual context diagram for software-testing education, training, and certification, that will help us better understand various "paths" via which a learner may go through, when learning testing. Section 4 discusses research questions and our experience-based research method for this paper. In Section 5, we present our observed states of software-testing education in academia and training in industry. In Section 6, we present competency profiling (modelling) of software-testing knowledge and skills. Finally, recommendations are drawn up in Section 7.

## 2 RELATED WORK

A number of papers have focused on using industry-academia collaborations for better (more effective) software-testing education and training.

A 2004 paper [3] studied the alignment of software testing skills of students with industry practices, from a South African perspective. The study identified significant differences between software testing skills required by industry and those learned by students.

A 2018 paper [4] proposed guidelines for improving software testing education using data from industry practices with a constructive alignment approach (an approach from education research [5]). The authors conducted a survey on how software organizations test their products, and used data to improve their testing courses. The principles of constructive alignment were used to present learning goals, teaching methods, and assessment methods that align with the industry requirements.

More broadly, the issue has also been discussed in engineering education as well. For instance, Arlett et al. [6] analyzed the relationships among academic staff, students and industry for experience-led engineering degree programs. The study made various recommendations, e.g., teaching by staff who have industry experience, input to the teaching from industry, and student's experience in industry.

## 3 RESEARCH DESIGN AND RESEARCH APPROACH

### 3.1 RESEARCH QUESTIONS

The two research question (RQs) that have directed our work on this paper are: (1) What is the state of the software-testing education, training and certification?; and (2) How can university educators and industrial trainers can cooperate to improve the state of education and training? We need to mention that comprehensive study of each RQ is multi-faceted and thus would require extensive studies, beyond this paper. However, in this paper, our intent is really to only "scratch the surface" on those two important issues (RQs), and to encourage further discussions in the community.

### 3.2 RESEARCH METHOD

The research method that we have used is participant-observation [7]. Both authors have had extensive experience in offering test education and training, in both academia and industry since the mid-2000's (see the next subsection). Via those years of experience, they have "observed" the state of the software testing education and training as "participants", and have earned the ability to suggest actionable recommendations.

### 3.3 AUTHORS' EXPERIENCE IN OFFERING TEST EDUCATION AND TRAININGS

The authors have had experience in offering test education and training, in both academia and industry since mid-2000's (Table 1).

The authors' experience covers teaching both classical university courses and also practical tool-focused training engagements in industry, e.g., how to use the test tools Selenium and Gauge (gauge.org). All the materials in this paper are based on those experiences. According to Table 1, the authors together have trained, so far, more than 3,000 learners in software testing.

**Table 1- Authors experience in test education and trainings**

| Author | University course or industrial training | Testing topics | # of learners |
|---|---|---|---|
| VG | Six offerings of an undergrad course on testing, U of Calgary, 2007-2012 | Overview of main testing topics with highly applied lab exercises [8] | Between 50-70 students each term, total ~ 360 |
| | Six offerings of a research-focused grad course on testing, U of Calgary, 2007-2012 | Research topics in testing with highly applied projects from industry [9] | Between 10-15 students each term, total ~ 60 |
| | Five offerings of an undergrad course on testing, Hacettepe U, Türkiye, 2013-2017 | Overview of main testing topics with highly applied lab exercises [8] | Between 70-90 students each term, total ~ 400 |
| | Five offerings of an undergrad course on testing, Queens Uni Belfast, UK, 2020-2024 | Overview of main testing topics with highly applied lab exercises [8] | Between 170-230 students each term, total ~ 1000 |
| | Customized corporate trainings to 85+ industry client companies since 2003 | Most have been focused topics, e.g., model-based testing, test process improvement using TMMi, or test automation tools, such as Selenium and Gauge | Sum of industrial learners ~ 350 |
| ABK | Corporate trainings to new hires and junior test engineers in the Testinium Corporation since 2018 | ISTQB Ffundamentals, test automation tools such as Selenium and Gauge | Sum of industrial learners ~ 500 |
| | Five offerings of an undergrad course on testing, İstanbul Kültür University and Bahçeşehir University, Türkiye, 2021-2024 | Overview of main testing topics with highly applied lab exercises, in collaboration with Testinium Corporation | Between 60-80 students each term, total ~ 350 |

## 4 CONTEXT OF SOFTWARE-TESTING EDUCATION, TRAINING AND CERTIFICATION

To better understand, characterize and distinguish between software-testing education in universities versus training in industry, we provide an abstract model, in the form of a context / process diagram, in Figure 1. On the left-hand side of the figure, there are universities that may or may not provide software testing education in their curriculum. We further distinguish between the SE/CS /IT degrees from non-SE, -CS, and -IT degrees, with the rationale that although graduates from the former types of degrees generally have higher knowledge in testing, and thus higher chance of getting employed as test engineers.

Educators in universities may decide to include or not include testing in their SE, CS, or IT curricula. As a result, in a world-wide view, we have graduates of SE, CS, and IT degrees with *varying* degrees of software testing skill-sets. After graduation, (most) university graduates look for positions in the software industry and are employed as SE professionals. A small ratio of graduates decide going to graduate school (MSc or PhD studies). A SE professional may or may not actively engage in software testing activities. We have also routinely observed that, in the software industry, many graduates of non-SE/CS/IT degrees (e.g., math, or business) also work in software testing positions. For example, in a survey of software testing practices in Canada in 2013 [10], based on a respondent population of 246 practitioners, 37% of all the respondents mentioned having non-SE/CS/IT degrees, e.g., business, MBA, industrial engineering, mathematics, and English.

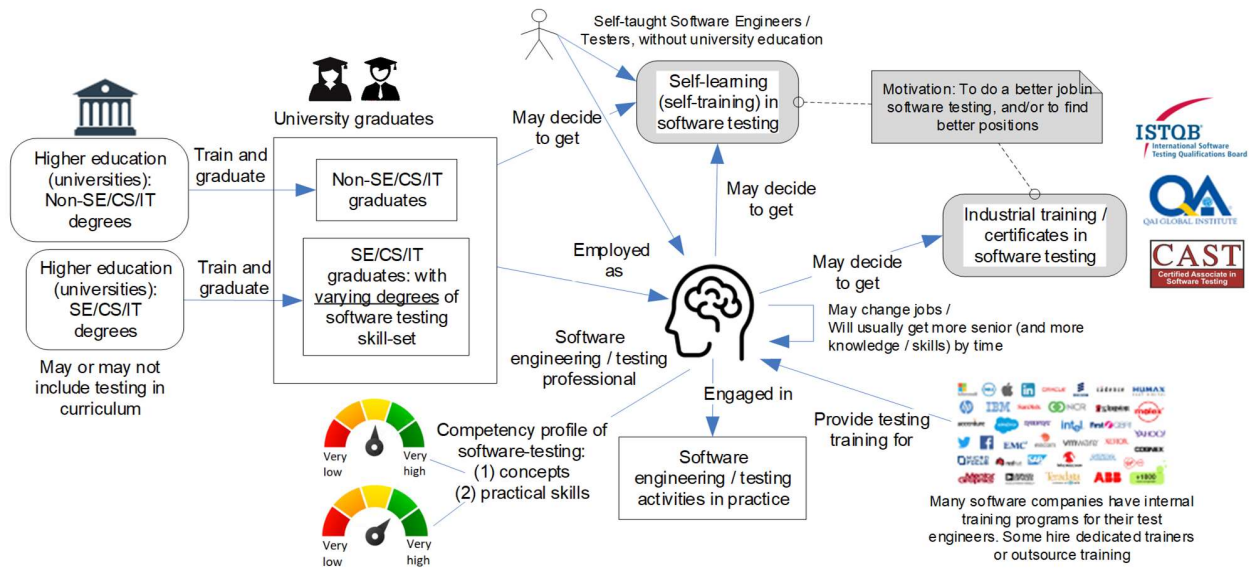

**Figure 1-A context diagram showing the relationship of software-testing education in universities versus training in industry**

Another important issue is that one does not (necessarily) need a university degree to become a software tester. A Google search for *become a software tester "without" university degree* returns more than 40 million search results (web pages and many videos). Although we were unable to find any global statistics about the ratio of practicing test engineers with versus without university degrees, in the above-discussed 2012 survey of software testing practices in Canada [10], 2% of the respondents reported having no university degrees.

To do a better job in software testing, and/or to find better positions, university graduates and practicing SE professionals may decide to self-learn software testing and/or self-improve their knowledge/skills using various types of resources, e.g., books and online learning resources and videos [11, 12]. They may also decide to attend various training courses (either in-person or online), and may also decide to get certificates in software testing, e.g., those provided by ISTQB, or the Certified Associate in Software Testing (CAST) issued by the International Software Certifications Board (ISCB), itself a subgroup within the Quality Assurance Institute (QAI).

ISTQB certifications seem to be the most popular among all certifications in testing, e.g., there are various opinions in the grey literature supporting it, such as the following: "*the range of progression paths following an ISTQB are much more varied and relevant for today's market*"[1]. The role of the established ISTQB certifications in the world of testing can be seen like the role of the TOEFL test for English language or the GRE (Graduate Record Examinations) test for graduate schools' admissions. Many companies explicitly mention the need for having an ISTQB certificate in their job postings.

Once a tester works in a testing position for a while, s/he can usually go up in a career path [13], often specified in the context of each company, e.g., from test analyst, to senior test analyst, and then test architect or test manager.

Furthermore, many software companies have internal training programs for their test engineers. The authors have observed this in the case of both SME (Small and Medium-sized Enterprise) software firms and large companies. For instance, as of 2008, Microsoft had a "SDET [Software Development Engineers in Test] Training Roadmap", spanning about 10 years in duration for its testers (see Figure 2-3 of [14]). Some companies hire dedicated trainers or outsource their training needs. For example, the first author has had years of experience offering dedicated and customized corporate training for a large number of companies. Note that all the corporate training materials must be as "applied" (hands-on) as possible, to ensure full engagement of learners. University-style training (with more focus on concepts and theory, rather than practice) often does not work in industrial training.

We should keep in mind that, in the SE domain, it is an accepted fact that, to be successful, software engineers need to be "life-long" learners [15]. Thus, a typical software tester can (should) expect to keep learning and going through the learning "flows" in Figure 1 for a long time (most of or all her/his career). For instance, although the first author has been developing test automation since 1998 and has used perhaps many different test automation tools already, when he enters a new testing project / team, and is asked to use a test automation tool, which is *new* to him, such as Playwright (playwright.dev) GUI automation tool in a recent project, he needs to put on his learner *hat* and learn the new skills, although he has been training testers for about two decades already.

---

[1] softwaretester.careers/which-software-testing-certification-is-best

Finally, let us discuss the two colorful gauges in Figure 1, which represent the two important components of software-testing competency profile of a given test engineer: testing knowledge (conceptual) and testing skills (practical). In essence, the entire efforts of various educational and training activities in Figure 1 is to increase these two competency aspects, for a given tester. We will discuss and focus on these two important items in Section 6.

## 5 Observations and challenges in software-testing education in academia and training in industry

Based on our combined experience, we discuss below a number of general observations and challenges in software-testing education and training, that we have observed over the years. Note that these observations and challenges are *general*, meaning that they may not apply in *all* contexts, but as per our experience of working in more than 10 academic and 100 industrial settings in six countries in three continents since 1998 and also as per our recent systematic mapping study [16], we believe that these are typical issues that are largely prevalent in academia and industry world-wide, more or less.

Software-testing education in academia and training in industry: Many CS / SE programs worldwide offer software-testing education in their courses [1]. The approach has been to either offer separate software testing courses [17], and/or integrate testing into programming courses [18].

- (-) /meaning negative observation: Some testing courses focus mostly on theoretical concepts and also their practical exercises are often rather small-scale [19]. Thus, such courses do not properly prepare students for the real-world large-scale testing. As a related issue, it has been reported that many students do not like learning too much about software testing [16] (many think that testing is tedious and boring, due to having too much theory). To address this issue, instructors can make those courses attractive and engaging (from students' viewpoint) by either making their course materials modern and large-scale like industrial context [8] or using industrial testing projects in courses [9].
- (-) Software-testing education in academia is largely non-uniform, i.e., contents coverage and depth can differ significantly from university to university and by the educator (who is teaching). Also, a major criticism in the industry [20] is about general inadequacy of software-testing education in university programs. Thus, when a graduated student approaches companies and states that s/he has taken the testing course in university, employers cannot be sure of what knowledge / skills s/he learnt. We have seen that, due to this challenge, many employers ask the student to take ISTQB certifications even if the student has reported that s/he has taken a testing course in his university studies already.
- (-) In terms of breadth and depth, testing education in university is generally much more limited compared to those in industry. For example, let's review different certifications of the ISTQB (Figure 2). Each certificate has an extensive curriculum, and based on our knowledge of ISTQB, it is fair to say that every ISTQB certificate has the knowledge and study load of roughly half a typical university course. Thus, given the limited learning capacity of typical university courses, we can only consider that testing education can and should only provide a *fundamental* base.

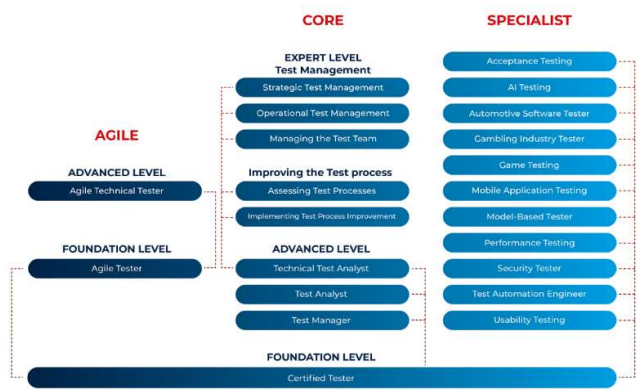

**Figure 2-Different certifications of the ISTQB**

Software-testing training and certification in industry:

- (-) Although some important testing concepts (such as test-case "design" techniques) are covered in most trainings, e.g., in the ISTQB foundation level, those techniques are not actively encouraged for usage, and thus many certified testers do not (properly) use those techniques, or only "satisfice" [21] that important aspect of testing in their projects. Note: to satisfice is to pursue a course of action that will satisfy the minimum requirements necessary to achieve a particular goal.
- (-) Due to the above mindset, most testing job postings and interviews largely focus on test "tools". When practitioners talk about testing, the discussions tend to mostly focus on which tools to use, but not "how" to properly use it, how to design test cases, etc.
- (+) / meaning positive observation: Compared to academia, contents coverage and depth is more uniform / standardized (especially for the ISTQB certificates). This often has many advantages, e.g., everyone can use the same vocabulary, and when a tester has gained a certificate, employers already know what knowledge s/he knows
- (+) ISTQB certifications have been established quite well in some countries, in the hiring process of many companies; and many employers / testers are already happy with the certification scheme, and its impact in the career path, etc.

Joint (academia and industry):

- (-) The two groups ("camps") prefer to continue operating (training testers) separately (it is a sad reality). Aside from educating and training testers, it is an unfortunate reality that, industry and academia do not cooperate extensively in other activities: neither in research [22], nor in conferences [23].
- (-) Due to non-uniform software-testing education in academia and highly fragmented streams of testers' entry to industry (see Figure 1), industry is providing its own training (ISTQB) from "scratch", as if university testing education does not exist

## 6 COMPETENCY PROFILING OF SOFTWARE-TESTING KNOWLEDGE AND SKILLS

Another important issue to consider is competency profiling of software-testing knowledge and skills. As per their definitions, knowledge is theoretical, whereas skills are practical [24, 25]. Competency profiling of software engineers is not only an active area of research [25], but also practice. There are various popular online tools such as CoderByte.com and TestDome.com that let companies to profile and evaluate competencies of job applicants and also their own employees quickly and accurately.

Two the models in this area are the Software Engineering Competency Model (SWECOM) [26], and the Software Engineering Body of Knowledge (SWEBOK) [27], both proposed by the IEEE. These models provide a number of knowledge areas and skills, respectively, for each area of SE, e.g., requirements, coding, testing and project management – see Table 2 and Figure 3 for software testing parts of those models. By reading each model's specification document, we can see that they have tried to cross reference the other document / model, but however, one aspect that we believe the two models fall short of, is the lack of strong linkage and cohesion between the models. In fact, in a recent test training activity in an industrial setting, the two authors tried to benefit from the two models to design and deliver a internal training initiative in Testinium A.Ş., but cross-referencing the items in the two models was not straightforward, as they have largely proposed their list of topics independently (see Table 2 and Figure 3 for the topics). One idea can be to merge (union) the two lists.

There are various approaches to gauge the level of knowledge and skills, e.g., the Individual Competency Index (ICI) [28] (see Figure 4), which is also used in other engineering fields. The five levels of the ICI index measure the depth of conceptual understanding and the extent of practical experience needed to perform an activity or task.

If we consider the two-dimensional space of levels of knowledge versus skills, we will get the diagram shown in Figure 5. We are also showing the general zones in which most university and industry testing courses fall in. University testing courses fall mostly in the right-bottom quadrant in which major emphasis is placed on theoretical aspects of testing, e.g., set theory for test-case design, etc. On other hand, most industry courses are applied and tool-based (e.g., how to use the Selenium framework), and often put (very) little emphasis on theoretical aspects. Cleary, it is almost impossible for any courses to cover "everything" in testing since training time is always limited. The ideal spot is always to learn as much theory as needed for the learning objective (e.g., if we are training our learners to conduct automated mobile app testing), and also to connect test tools usage to fundamental concepts, e.g., test-case design. The open-source lab exercises that we have developed since 2010 [8] have had this goal in mind. Note that open-source courseware has been used by 100+ educators so far.

**Table 2- Software testing competencies of SWECOM (from Section 14 of [26])**

| Software testing skills | Software testing activities |
|---|---|
| Software test planning | • Identify all stakeholders involved in software testing<br>• Identify success and failure criteria<br>• Identify test completion criteria<br>• Design and implement the software test plan<br>• Identify and coordinate customer representatives and other stakeholders participating in the software acceptance and/ or demonstration |
| Software testing infrastructure | • Identify tools to be used throughout testing activities<br>• Identify appropriate documentation to be generated and archived<br>• Design/select and implement the test environment |
| Software testing techniques | • Identify test objectives<br>• Select appropriate testing/demonstration techniques<br>• Design, implement, and execute test cases |
| Software testing measurement and defect tracking | • Identify, collect, and store appropriate data resulting from testing/demonstration<br>• Report test results to appropriate stakeholders<br>• Identify, assign, and perform necessary corrective actions<br>• Analyze test data for test coverage, test effectiveness, and process improvement |

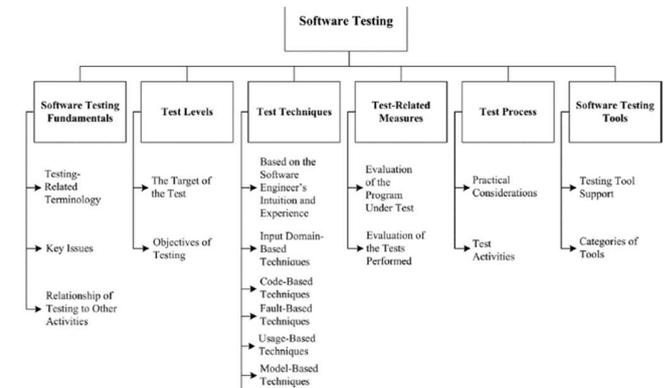

**Figure 3-Topics of the software testing knowledge area, as suggested by SWEBOK [27]**

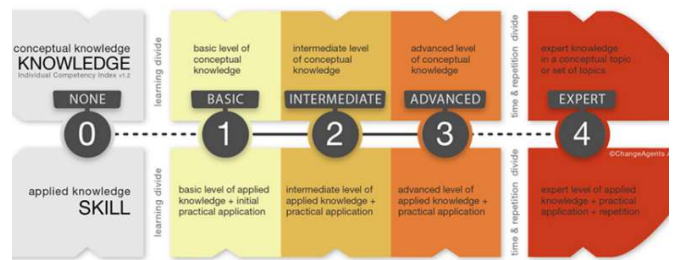

**Figure 4-Five levels of the Individual Competency Index (ICI) [28]**

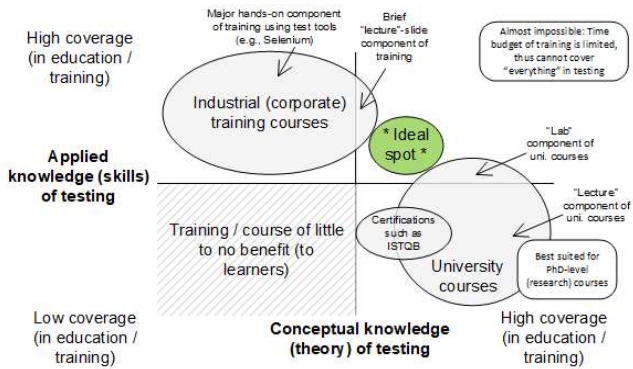

**Figure 5-Levels of knowledge versus skills**

## 7 RECOMMENDATIONS

Based on all the above discussions, we provide a number of recommendations.

<u>Right mix of theory-practice:</u>

A given education / training course should have the "right" mix of theory-practice for its learners. Even for university students, too much theory (e.g., too much focus on formal methods in testing) can bore the students and decrease their interest in our field, as they will struggle to see how such theoretical materials can be used in practice.

The authors' teaching philosophy in the past 20 years [8, 9] has taken such as approach and they have constantly received excellent feedbacks from learners, both in university and industry. Even if when we have been teaching tool-focused industry trainings (such as Selenium and Gauge BDD tool), we have included some theoretical test-case design techniques (e.g., multi-dimensional equivalence classing) and have immediately applied them in the training sessions to test a real software, to ensure learners properly learn the concept.

<u>Roles of academia and industry:</u>

The current state of testing education and training can only be improved if educators / trainers from both sides take proactive steps; and if they come out of their "comfort zones". The authors have been working hard to do that in the past two decades.

Symbiotic collaboration of academia and industry in this endeavor to ensure effective education of young test engineers who have learned some foundation of pragmatic testing theory, plus skills on how to conduct different types of testing using modern test tools. Two example ideas that we have been doing in the past two decades are: (1) incorporating real-world industrial testing projects in software testing courses [9], and (2) designing practical (lab) extensive in which real-world test tools (such as Junit and Selenium) are used by students to test non-trivial (large) software systems under test [8]. With such efforts, we can move towards closing the gap between software testing education and industrial needs [29].

## 8 CONCLUSIONS

While it is clearly the case that a SE, CS, or IT degree trains a student to have important foundational skill-set of testing in practice, many believe [20] that graduates often also needs additional testing training after university, e.g., via certifications schemes of ISTQB. This paper aimed at providing a pragmatic overview of the issue and to trigger further discussions in the community, and between industry and academia, on how to further improve the status quo and also find further best practices with cooperation of industry and academia in education and training of software testers. As Mahatma Gandhi said: "*Be the change you wish to see in the world*".